\documentclass[10pt,a4paper]{article}

\usepackage[english]{babel}
\usepackage[utf8]{inputenc}
\usepackage[T1]{fontenc}
\usepackage{graphicx}
\usepackage{geometry}
\usepackage{tabularx}
\usepackage{psfrag}
\usepackage{fancyhdr}
\usepackage{xcolor}
\usepackage{sistyle}
\usepackage{amsfonts}
\usepackage{amsmath}
\usepackage{amssymb}
\usepackage{booktabs}
\usepackage{multirow}
\usepackage{amsthm}
\usepackage{subcaption}
\usepackage{pifont}
\usepackage{enumerate}
\usepackage{enumitem}
\usepackage[numbers, sort&compress]{natbib}
\usepackage{bbold}
\usepackage{bbm}
\usepackage{wrapfig}

\usepackage{amssymb,amsfonts}

\usepackage{nicefrac}

\usepackage{subcaption}

\usepackage{algorithm} 
\usepackage{algorithmic}

\usepackage{booktabs}
\usepackage{multirow}

\usepackage{footmisc}

\usepackage{xcolor}

\usepackage[colorlinks]{hyperref}

\usepackage{placeins}

\newlength{\myeqskip}  
\def\eg{e.g.}

\def\wdkl{\omega}

\def\dkl{\mathsf{d}_{\mathsf{KL}}}
\def\Dkl{\mathsf{D}_{\mathsf{KL}}}

\def\Z{\mathsf{Z}}
\def\Zvect{\boldsymbol{\mathsf{Z}}}

\def\P{\mathsf{P}}

\def\Zphi{\Phi^{\Zvect}}
\def\Zphivect{\boldsymbol{\Phi}^{\Zvect}}

\def\R{\mathsf{R}}
\def\P{\mathsf{P}}
\def\Rvect{\boldsymbol{\mathsf{R}}}

\def\lambdatime{\lambda_{\mathsf{T}}}
\def\lambdaspace{\lambda_{\mathsf{S}}}
\def\lambdafro{\lambda_{\mathsf{L}}}

\def\B{\mathsf{B}}
\def\D2{\mathsf{D}_2}
\def\K{\mathsf{K}}
\def\L{\mathsf{L}}
\def\M{\mathsf{M}}
\def\tr{\operatorname{Trace}}

\def\Qvect{\boldsymbol{\mathsf{Q}}}

\def\argmin{\mathrm{argmin}}
\def\F{F} 
\def\G{G} 
\def\J{J} 
\def\italt{n}

\def\rset{\mathbb{R}}

\def\admissibleR{\mathcal{R}}
\def\admissibleL{\mathcal{L}}

\def\W{\mathsf{W}}
\def\Fro{\operatorname{Fro}}

\def\mRSE{\mathsf{mRSE}}
\def\mRSE{\mathsf{mRSE}}

\def\ML{\mathsf{ML}}
\def\Epi{\mathsf{EpiEstim}}
\def\fixL{\mathsf{fix-}\L}
\def\fixR{\mathsf{fix-}\R}
\def\joint{\mathsf{Joint}}

\graphicspath{{figures/}}

\title{\Large Joint Reproduction Number and Spatial Connectivity Structure Estimation via Graph Sparsity-Promoting Penalized Functional}
\author{\large Etienne Lasalle$^{1}$, Barbara Pascal$^{1}$\footnote{B. Pascal is partly supported by ANR-23-CE48-0009 ``\textit{OptiMoCSI}".}}
\date{\normalsize $^{1}$ Nantes Universit\'e, \'Ecole Centrale Nantes, CNRS, LS2N, UMR 6004, F-44000 Nantes, France, \, \texttt{firstname.lastname@ls2n.fr}}

\begin{document}

\maketitle

\noindent{\bf Abstract.}
During an epidemic outbreak, decision makers crucially need accurate and robust tools to monitor the pathogen propagation.
The effective reproduction number, defined as the expected number of secondary infections stemming from one contaminated individual, is a state-of-the-art indicator quantifying the epidemic intensity.
Numerous estimators have been developed to precisely track the reproduction number temporal evolution.
Yet, COVID-19 pandemic surveillance raised unprecedented challenges due to the poor quality of worldwide reported infection counts. 
When monitoring the epidemic in different territories simultaneously, leveraging the spatial structure of data significantly enhances both the accuracy and robustness of reproduction number estimates.
However, this requires a good estimate of the spatial structure. 
To tackle this major limitation, the present work proposes a joint estimator of the reproduction number and connectivity structure.
The procedure is assessed through intensive numerical simulations on carefully designed synthetic data and illustrated on real COVID-19 spatiotemporal infection counts.

\medskip


\section{Introduction}
\label{sec:intro}

\noindent{\bf Context.}
Accurate and efficient monitoring tools are of key importance during epidemic outbreaks.
The effective reproduction number, a widely used indicator, captures the daily evolution of the intensity of an epidemic by quantifying the secondary infections stemming from a typical infectious individual~\cite{Diekmann1990,wallinga2004,cori2013new,Liu2018,thompson2019improved}.
A major challenge in real-time pandemic surveillance lies in the potentially very low quality of reported infection counts,  showing pseudo-seasonalities,  underestimation on days-off,  delays~\cite{abry2020spatial,pascal2022nonsmooth}. 
State-of-the-art reproduction number estimators, although enforcing a mild temporal regularity of the estimate~\cite{cori2013new,thompson2019improved}, are not robust enough to handle such low data quality.
This motivated the design of more advanced \emph{variational} strategies using sparsity-based regularizations to yield more realistic and readable estimates~\cite{abry2020spatial,pascal2022nonsmooth}.
Furthermore, as pandemics are simultaneously monitored in a large number of territories (e.g., different countries, counties of a state, states of a federal country), extensions of these variational estimators took advantage of the \emph{spatial} structure of available data to enhance the accuracy of the reproduction number estimates via a graph-based spatial regularization~\cite{abry2020spatial,pascal2022nonsmooth,du2024synthetic}.
Numerical simulations on synthetic data demonstrated that taking into account the spatial structure significantly improves the estimate quality, provided that the connectivity structure between territories is \emph{known}~\cite{du2024synthetic}.
In practice, this structure is inaccessible and, while using the unweighted vicinity graph, in which territories sharing terrestrial borders are connected,  showed promising results on French \emph{départements}\cite{abry2020spatial,pascal2022nonsmooth}, it is too simplistic to analyze the pandemic at larger scales.
To that aim, data-driven procedures inferring the connectivity structure are crucially needed.

\noindent{\bf Related works.} 
Graph learning is a long-standing problem taking roots not only in graph processing but also in statistics, notably leading to the very popular graphical lasso estimator~\cite{dempster1972covariance,friedman2008sparse}, in machine learning, for semi-supervised learning~\cite{ghanem2024gradient}, and network analysis~\cite{kolaczyk2014statistical}.
However, it only recently emerged as a topic of interest in graph signal processing~\cite{dong2019learning,mateos2019connecting}.
In this framework, a popular strategy is to learn a graph in the topology of which the signal behaves smoothly,  where the smoothness can be for example enforced via Laplacian-based total variation~\cite{shuman2013emerging} which has some connections with graphical lasso~\cite{dong2019learning}.
During the past decade, joint estimation of the underlying graph and of the signal, in graph signal processing, or labels, in machine learning,  received much attention~\cite{dong2016learning, dong2019learning, ghanem2024gradient}. 
A major advantage of joint strategies is that they result in a significantly smaller estimation error than two-step procedures, estimating first the graph from noisy data and then the regularized signal.

\noindent{\bf Goals, contributions and outline.} 
Given infection count time series in several territories, the overarching goal of the present work is to \emph{jointly} learn the spatial connectivity structure and estimate the reproduction number along time and across territories in an unsupervised manner.
To that aim, elaborating on the graph representation framework developed in~\cite{dong2016learning}, the spatial structure is encoded in a weighted graph, represented through its combinatorial \emph{Laplacian} matrix.
A major originality of the present work is to perform Laplacian estimation under Poisson data-dependent noise, contrary to most of previous Laplacian estimation procedures developed under Gaussianity assumption~\cite{dong2016learning,egilmez2017graph}.
The second major originality is that, contrary to standard graph signal processing frameworks in which the signal is \emph{real}-valued (e.g., year of graduation in a social network, transportation capacities at hub in a transportation network), each territory has its own infection count \emph{time series}.
Section~\ref{sec:model} recasts the univariate temporal epidemiological model~\cite{cori2013new}, into a multivariate spatiotemporal setting following~\cite{abry2020spatial,pascal2022nonsmooth}, leverages the \emph{scaled Poisson} model~\cite{pascal2024risk} to enable variance modulation across territories, and adapts the graph-regularized reproduction number estimator from~\cite{abry2020spatial} to this novel setting.
The proposed joint estimation of the spatiotemporal reproduction number and connectivity structure is developed in Section~\ref{sec:graph}.
Section~\ref{sec:numerics} demonstrates on realistic synthetic data both the significant estimation accuracy gain obtained leveraging the spatial structure and the ability of the proposed method to retrieve very accurately the underlying connectivity structure and illustrates the method on real COVID-19 infection counts.
Concluding remarks and perspectives are provided in Section~\ref{sec:conclusion}.

\noindent{\bf Notations.} Real (resp. non-negative real) numbers are denoted by $\rset$ (resp. $\rset_+$) and positive integers by $\mathbb{N}^*$.  
Scalars (resp. time series) are written in plain (resp. bold) characters
and linear operators in upper case sans serif font. 
The component-wise product (resp. division) is denoted by $\odot$ (resp. $\centerdot /$).
Ground truth quantities are stared.

\section{Spatiotemporal epidemiological model} 
\label{sec:model}

\begin{figure}[t!]
\includegraphics[width=0.325\linewidth]{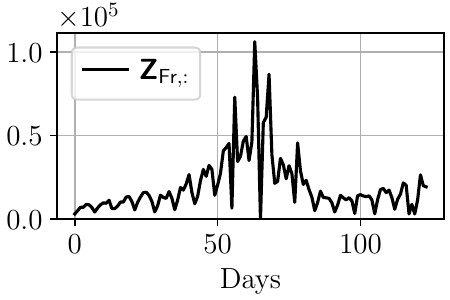}
\includegraphics[width=0.325\linewidth]{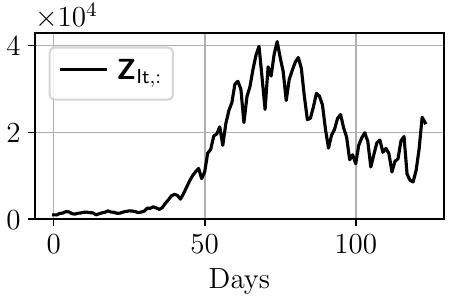}
\includegraphics[width=0.325\linewidth]{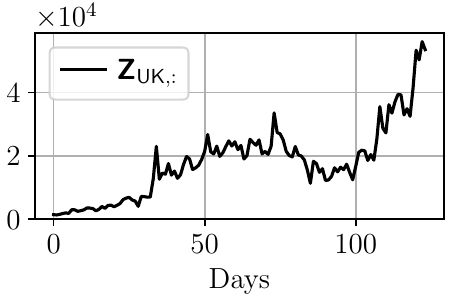}
\caption{\textbf{COVID-19 infection counts} reported in France, Italy, and the United-Kingdom from Sept.  $1$, $2020$ to Oct. $1$, $2021$.
\footref{ft:jhu}}
\label{fig:trueZs}
\end{figure}

\noindent \textbf{Scaled Poisson spatiotemporal epidemiological model.}
The very small variance of the Poisson epidemiological model proposed in~\cite{cori2013new} poorly accounts for the large variability of COVID-19 data illustrated in Fig.~\ref{fig:trueZs} displaying infection counts in France, Italy and United-Kingdom collected from National Health Authorities and made available by the Johns Hopkins University.\footnote{\label{ft:jhu}\href{https://coronavirus.jhu.edu/}{coronavirus.jhu.edu}} 
To better model COVID-19 propagation, a \emph{scaled} Poisson model, with an extra scale parameter controlling the infection counts variance, has been proposed in~\cite{pascal2024risk}.
The first contribution of this work is to extend this univariate description into a \emph{multivariate} model of infection counts in different territories, hereafter referred to as \emph{countries}.
In country $c$ at time $t$ the infection count $\Z_{c,t}$, conditionally to past counts $\Z_{c,1},\hdots,\Z_{c,t-1}$, is modeled as a scaled Poisson random variable
\begin{align}
\label{eq:multivar-Z}
\frac{\Z_{c,t}}{\gamma_c} \lvert  \Z_{c,1}, \hdots, \Z_{c,t-1} ; \R_{c,t} \sim \mathcal{P}\left( \frac{\R_{c,t} \Zphi_{c,t}}{\gamma_c}\right).
\end{align}

The spatiotemporal Poisson parameter combines the effective reproduction number $\R_{c,t}\geq 0$, the \emph{global infectiousness} $\Zphi_{c,t} = \sum_{s = 1}^{\tau_{\Phi}} \phi(s) \Z_{c,t-s} $, which is a weighted sum of past counts where the serial interval function $\phi$ accounts for the random delay between primary and secondary infections, and a scale parameter $\gamma_c >0$,  constant over time,  modulating count variance in country $c$. 
For COVID-19 pandemic, $\phi$ is modeled by a gamma distribution of mean $6.6$~days and standard deviation $3.5$~days truncated at $\tau_{\Phi} = 25$ days~\cite{guzzetta2020,riccardo2020epidemiological}.
The infection count time series over $T$ days in the $C$ different countries are represented as column vectors $\Zvect_{c,:} \in \mathbb{R}^{T}$ for $c \in \lbrace 1, \hdots, C\rbrace$; the concatenation of the $C$ time series is encoded in matrix form $\Zvect \in \mathbb{R}_+^{C \times T}$.
The present model assumes that the $C$ random times series $\Zvect_{c,:}, \, c \in \lbrace 1, \hdots, C\rbrace$ are \emph{independent} and thus, that pandemic dynamic similarities between countries are \emph{entirely} encoded in the resemblance of their reproduction number profiles.

\noindent \textbf{Regularized spatiotemporal variational estimator.} 
Due to the low quality of reported counts illustrated in Fig.~\ref{fig:trueZs}, the reproduction number estimation requires advanced regularizing strategies.
A first building block of the present work consists in designing a variational estimator of spatiotemporal reproduction numbers adapted to the multivariate scaled Poisson model of Eq.~\eqref{eq:multivar-Z}  enforcing a  realistic, hence smooth, behavior both in time and across \emph{connected} countries given a \emph{fixed} connectivity structure.
Following~\cite{abry2020spatial,pascal2022nonsmooth,du2024synthetic,pascal2024risk}, the data-fidelity term of~\cite[Eq.~(7)]{abry2020spatial} is replaced by a multivariate scaled Kullback-Leibler divergence, designed from the negative log-likelihood of the scaled Poisson model derived in~\cite[Eq.~(20)]{pascal2024risk}.
Temporal regularity is obtained by enforcing a piecewise linear temporal behavior of the reproduction number, independently on each country, via the penalization of the $\ell_1$-norm of its time second derivative. 
To favor a smooth spatial behavior, the connectivity structure is encoded in a graph whose vertices are the countries and with edges linking countries expected to share similar epidemiological dynamics and a graph total variation penalization~\cite{shuman2013emerging} is used to regularize the spatial variations of the reproduction number.
While~\cite{abry2020spatial} used an unweighted graph, the present work proposes to consider a \emph{weighted} graph, enabling to account for different similarity levels and hence for more intricate, and thus more realistic, connectivity structures.
Finally, to ease the design of a Laplacian learning strategy in Section~\ref{sec:graph}, the $\ell_1$ spatial regularization is turned into a squared $\ell_2$ term.
Altogether, this leads to the customized variational estimator
\begin{align}
\label{eq:argmin_R}
\widehat{\Rvect}^{\fixL} =\underset{\Rvect \in \mathbb{R}_+^{C\times T}}{\argmin} &\sum_{c = 1}^C \bigl( \wdkl_c \Dkl\bigl(\Zvect_{c,:} \left\lvert \Rvect_{c,:} \odot \Zphivect_{c,:}\right. \bigr)  
\!+\! \lambdatime  \lVert \D2 \Rvect_{c,:} \rVert_1\bigr) \!+\! \lambdaspace \sum_{t = 1}^T \lVert \B \Rvect_{:,t} \rVert_2^2 
\end{align}
where $\Dkl\bigl(\Zvect_{c,:} \lvert \Rvect_{c,:} \odot \Zphivect_{c,:} \bigr)= \sum_{t=1}^T \dkl\bigl(\Z_{c,t} \lvert \R_{c,t}  \Zphi_{c,t}\bigr) $ with $\dkl$ the Kullback-Leibler divergence;\footnote{The Kullback-Leibler divergence is defined as
$ \mathsf{d}_{\mathsf{KL}}(\Z\lvert \P) =  \Z \ln\left( \Z/\P\right) + \P - \Z, \,  \text{if } \, \Z>0, \, \P>0; 
 \P, \,  \text{if }\, \Z=0, \, \P \geq 0; \infty,  \, \text{otherwise.}$
}  the weights $\wdkl_c > 0$ in the data-fidelity term would ideally be $\wdkl_c = 1/\gamma_c$ but could be replaced by heuristics when the scale parameters are unknown; the matrix $\D2 \in \mathbb{R}^{T-2 \times T}$ denotes the second-order difference operator; $\B$ is the incidence matrix\footnote{Given an undirected graph with $C$ vertices and $E$ edges, the incidence matrix $\B \in \mathbb{R}^{E\times C}$ is constructed so that if $e = \lbrace c, c'\rbrace$ is an arbitrarily oriented edge between vertices $c$ and $c'$ with $c < c'$, then $\B_{e,c} = - \B_{e,c'} = \sqrt{\W_{c,c'}}$ and $\B_{e, c''} = 0$, $\forall c'' \notin \lbrace c, c'\rbrace$, where $\W_{c,c'}$ is the weight associated to edge $\lbrace c, c'\rbrace$ with $\W_{c,c'}\in \lbrace 0,1\rbrace$ for unweighted graphs.} of the \emph{fixed} graph encoding connectivity between countries; and $\lambdatime > 0$ (resp. $\lambdaspace>0$) are hyperparameters controlling the level of temporal (resp. spatial) regularization.

The usual definition of the graph total variation involves the \emph{Laplacian} matrix of the graph, denoted by $\L = \B^\top \B$.
Remarking that $\lVert \B \Rvect_{:,t} \rVert_2^2 = \Rvect_{:,t}^\top \L \Rvect_{:,t}$ shows that the graph-based penalization in~\eqref{eq:argmin_R} is a rewritting of the graph total variation.
Furthermore,  replacing the incidence matrix $\B\in \mathbb{R}^{E\times C}$ by any matrix $\widetilde{\B} \in \mathbb{R}^{C \times C}$ satisfying $\widetilde{\B}^\top \widetilde{\B}= \L$ 
lets the graph total variation unchanged, while significantly improving computational and memory efficiency when the number of edges $E$ is far larger than the number of countries $C$.
The objective function in~\eqref{eq:argmin_R} is lower semi-continuous, proper, lower bounded by zero and coercive. 
Hence there exists a solution to \eqref{eq:argmin_R}.
As the sum of convex functions composed with linear operators, the objective function is convex.
Furthermore, the Kullback-Leibler divergence is strictly convex with respect to its second argument, then provided that $\Zphi_{c,t }> 0$ for all countries $c$ and times $t$, the solution of~\eqref{eq:argmin_R} is unique, and hence the $\fixL$ estimator is well-defined.

\noindent \textbf{Algorithmic resolution.}
Due to the presence of the $\ell_1$-norm, the objective function is nonsmooth, discarding the use of gradient-based minimization algorithms.
Instead one has to resort to \emph{proximal} algorithms~\cite{parikh2014proximal}. 
The objective function in~\eqref{eq:argmin_R} can be rewritten as the sum of functions with closed-formed 
proximity operators~\cite{prox_repository} possibly composed with linear operators: $\F(\Rvect) + \G(\K \Rvect)$
with $\displaystyle \F(\Rvect) = \sum_{c = 1}^C \wdkl_c \Dkl\left(\Zvect_{c,:} \left\lvert \Rvect_{c,:} \odot \Zphivect_{c,:}\right. \right) $, $\K$ combining the linear operators involved in the regularization terms of~\eqref{eq:argmin_R} such that,  $\K \Rvect =$
\begin{align}
\label{eq:def_K}
 \bigl(  \D2 \Rvect_{1,:},\hdots,\D2 \Rvect_{C,:}, \B\Rvect_{:,1}, \hdots,\B\Rvect_{:,T}\bigr) = (\Qvect_{\mathsf{T}},\Qvect_{\mathsf{S}}) = \Qvect
\end{align}
where $\B$ is potentially equivalently replaced by $\widetilde{\B}$, and $G$ defined on the range of $\K$ as $\displaystyle \G(\Qvect_{\mathsf{T}},\Qvect_{\mathsf{S}}) = \lambdatime \lVert \Qvect_{\mathsf{T}}  \rVert_1 + \lambdaspace \lVert \Qvect_{\mathsf{S}}  \rVert_2^2$, where for $p \in \lbrace 1, 2 \rbrace$,  $\lVert \cdot \rVert_p^p$ applies in a \emph{componentwise} manner.
The functions $\F, \G$ being proper, lower-semicontinous, convex,  following\cite[Algorithm~1]{pascal2022nonsmooth}, the minimization problem~\eqref{eq:argmin_R} can be solved taking advantage of the Chambolle-Pock primal-dual scheme~\cite{chambolle2011first}. 
Furthermore,  the robust criterion proposed in~\cite[Eq.~(8)]{du2023compared} is leveraged to stop the iterations if a predefined precision $\varepsilon > 0$ is reached
or if the maximum budget of iterations $k_{\max}$ has been exhausted.

\section{Joint estimation strategy}
\label{sec:graph}

\begin{algorithm}[t!]
\caption{Alternating optimization}\label{alg:AO}
\begin{algorithmic}[1]
{\small\REQUIRE $\Zvect \in \mathbb{R}^{T}$, $(\lambdatime, \lambdaspace, \lambdafro) \in \mathbb{R}_+^3$, 
$\italt_{\max} \in \mathbb{N}^*$, $\Rvect^{(0)}$; \\
\STATE $\L^{(0)}, \M^{(0)} \gets \texttt{minimizeL}\bigl(\Rvect^{(0)}, \lambdaspace, \lambdafro \bigr)$
\STATE $\widetilde{\B}^{(0)} \gets \texttt{Cholesky}(\L^{(0)})$; define $\K$ from~\eqref{eq:def_K} using $\widetilde{\B}^{(0)}$
\STATE $\Qvect^{(0)} \gets \K \Rvect^{(0)}$

\hspace*{-2.05em} \textbf{Alternating optimization:}
\FOR{$\italt = 0$ \TO $\italt_{\max}-1$}
    \STATE $\Rvect^{(\italt+1)}, \Qvect^{(\italt+1)} \gets \texttt{minimizeR}\bigl(\Zvect, \lambdatime, \lambdaspace, \widetilde{\B}^{(\italt)}, \Rvect^{(\italt)}, \Qvect^{(\italt)}\bigr)$
    
    \vspace{-0.5mm}
    \STATE $\L^{(\italt+1)}, \M^{(\italt+1)} \gets \texttt{minimizeL}\bigl(\Rvect^{(\italt+1)}, \lambdaspace, \lambdafro, \L^{(\italt)}, \M^{(\italt)}\bigr)$
        
    \vspace{-0.5mm}
    \STATE $\widetilde{\B}^{(\italt+1)} \gets \texttt{Cholesky}(\L^{(\italt+1)})$
\ENDFOR

\RETURN number $\Rvect^{(\italt_{\max})}, \L^{(\italt_{\max})}$}
\end{algorithmic}
\end{algorithm}

Given a multivariate count time series $\Zvect \in \rset^{C \times T}$, the overarching goal of the present work is to estimate \emph{jointly} the reproduction number $\Rvect$ and the underlying connectivity structure. 
Among the numerous methods proposed in both graph signal processing and machine learning to address the simultaneous estimation of graph signals and connectivity structure, three are adapted to the variational framework: bi-level optimization~\cite{ghanem2024gradient}, 
and joint~\cite{fatemi2021slaps} and alternate~\cite{dong2016learning} minimization. 
Both bi-level and joint optimization require strong regularity and convexity properties on the respective objective functions involved to get convergence guarantees, thus, following~\cite{dong2016learning}, alternating minimization has been preferred.
Two major contributions of the present work consist in designing a \emph{joint} objective function 
and deriving a rapid alternating minimization scheme.

\noindent \textbf{Regularized and sparsity-inducing objective function.}  Sticking to the variational framework of Section~\ref{sec:model}, the joint estimation of $\Rvect$ and $\L$ is formulated as a constrained optimization problem:
\begin{equation}
    \widehat{\Rvect}^{\joint} , \widehat{\L}^{\joint} \in \underset{\Rvect \in \admissibleR, \L \in \admissibleL}{\mathrm{Argmin}} \, \J(\Rvect, \L)\, .
    \label{eq:joint_optim_pb}
\end{equation}
The smart design of the objective function $J$ and admissible sets $\admissibleR$, $\admissibleL$ enables to enforce simultaneously spatial and temporal regularity of $\Rvect$, so that estimates are epidemiologically realistic, and sparsity of $\L$, so that the graph only contains the most salient connections.
Combining the regularized estimate 
defined in~\eqref{eq:argmin_R} and the graph learning strategy adapted to scalar Gaussian graph signals from~\cite{dong2016learning} yields
\begin{align}
\begin{split}
\label{eq:full_obj_function}
\J(\Rvect, &\L) = \sum_{c = 1}^C \wdkl_c \Dkl\left(\Zvect_{c,:} \left\lvert \Rvect_{c,:} \odot \Zphivect_{c,:}\right. \right) + \lambdatime  \sum_{c = 1}^C  \lVert \D2 \Rvect_{c,:} \rVert_1  + \lambdaspace \sum_{t = 1}^T \Rvect_{:,t}^\top \L \Rvect_{:,t} + \lambdafro \lVert \L \rVert_{\Fro}^2 \,, 
\end{split}
\end{align}
where the graph total variation has been written using the Laplacian matrix instead of the incidence matrix as in~\eqref{eq:argmin_R}, $\lVert \cdot \rVert_{\Fro}$ denotes the Frobenius norm, $\lVert \L \rVert_{\Fro}^2 = \sum_{c,c'} \L_{c,c'}^2$,
and  $\lambdafro > 0$ controls the estimated $\widehat{\L}^{\joint}$ sparsity level.
The admissible sets are chosen as 
\begin{align}
\label{eq:admissible}
\begin{split}
\admissibleR  &= \{ \Rvect \in \rset^{C \times T}, \ \text{s.t. } \forall c, t \ \Rvect_{c,t} \geq 0  \} \,,\\
\admissibleL &= \left\{ \L \in \rset^{C \times C}, \ \text{s.t. } \tr(\L)=C, \right. \left. \forall c \neq c' \  \L_{c, c'} = \L_{c',c} \leq 0 \text{ and } \L \boldsymbol{1}_C = \boldsymbol{0}_C  \right\} \,,
\end{split}
\end{align}
where $\boldsymbol{0}_C$ (resp. $\boldsymbol{1}_C$) is the vector of size $C$ with constant value 0 (resp. 1). 
To be consistent with~\eqref{eq:multivar-Z},  $\admissibleR$ enforces nonnegativity of the estimated reproduction numbers, while $\admissibleL$ ensures the estimation of a proper Laplacian matrix with a fixed total weight, excluding the trivial empty graph solution.
The smaller the regularization parameter $\lambdafro$, the less the total weight is able to spread,  hence the sparser the estimated graph.
The objective function $J$ is continuous, lower bounded by zero, coercive in $\Rvect$ and defined on the compact admissible set $\admissibleL$ for $\L$, thus the minimization problem~\eqref{eq:joint_optim_pb} has at least one solution. 
Yet, the objective function, while convex in $\Rvect$ and $\L$ separately, is not jointly convex.
Minimizing~\eqref{eq:full_obj_function} thus requires to resort to an alternating scheme, successively minimizing $J$ with respect to $\Rvect$ (resp. $\L$) at fixed $\L$ (resp. $\Rvect$).
The designed scheme is detailed in Algorithm~\ref{alg:AO}.
The inner optimization procedures are both solved efficiently with primal-dual schemes and systematically initialized using the previous primal and dual solutions to speed up convergence.

\noindent \textbf{Minimization with respect to $\Rvect$ at fixed $\L$.}
By equivalence of the incidence and Laplacian matrices-based formulations of graph total variation, the terms in $J$ which depends on the reproduction number exactly match the objective function in~\eqref{eq:argmin_R}.
Thus, the inner step of Algorithm~\ref{alg:AO} minimizing over $\Rvect$ can be performed leveraging the Chambolle-Pock scheme customized to~\eqref{eq:argmin_R} with connectivity structure encoded in a matrix $\widetilde{\B}$ link to the \emph{fixed} Laplacian through $\L = \widetilde{\B}^\top\widetilde{\B}$.
For the sake of computational efficiency, $\widetilde{\B}$ is chosen of size $C\times C$ and obtained from Cholesky decomposition of $\L$.\footnote{\href{https://numpy.org/doc/stable/reference/generated/numpy.linalg.cholesky.html}{numpy.org/doc/stable/reference/generated/numpy.linalg.cholesky}}
At first call in Algorithm~\ref{alg:AO}, the primal and dual variables of the Chambolle-Pock algorithm are initialized at $\Rvect^{[0]} = \Zvect \centerdot/\Zphivect$ and $\Qvect^{[0]} = \K \Rvect^{[0]}$.

\noindent \textbf{Minimizing with respect to $\L$ at fixed $\Rvect$.}
The terms of $J$ depending on $\L$, together with the constraints encoded in $\admissibleL$, rewrites
\begin{align}
    \widehat{\L}^{\fixR} = & \underset{\L}{\mathrm{argmin}} \ \lambdafro \sum\limits_{c,c'=1}^C \L_{c,c'}^2 +  \lambdaspace \sum\limits_{c,c'=1}^C  (\Rvect_{c,:}^\top \Rvect_{c',:}) \L_{c,c'}  \,, \label{eq:argmin_L} \\
    \text{s.t. } & \tr(\L) = C \,, \L_{c, c'} = \L_{c', c} \leq 0 , \ \forall c \neq c' \,, 
     \L \boldsymbol{1}_C = \boldsymbol{0}_C \,. \nonumber
\end{align}
Since $\lambdafro >0$, the objective function in~\eqref{eq:argmin_L}, defined on a convex domain,  is non-negative, continuous, coercive,  and strictly convex, therefore it admits a unique minimizer.
Furthermore,  remarking that the constrained minimization problem~\eqref{eq:argmin_L} is a quadratic program subjected to linear constraints enables to solve it very efficiently using an interior point method~\cite{grant2008graph,andersen2011interior}.
In practice, the resolution of~\eqref{eq:argmin_L} is performed by the quadratic program solver\footnote{\href{https://cvxopt.org/userguide/coneprog.html\#quadratic-programming}{cvxopt.org/userguide/coneprog.html\#quadratic-programming}} of the CVXOPT package~\cite{andersen2013cvxopt}.
At first call in Algorithm~\ref{alg:AO},  the primal and dual variables of the quadratic program are set to the solver default values.

\section{Numerical experiments}
\label{sec:numerics}

\begin{figure}[t!]
	\centering
    \begin{minipage}{0.44\linewidth}
    \includegraphics[width=\linewidth]{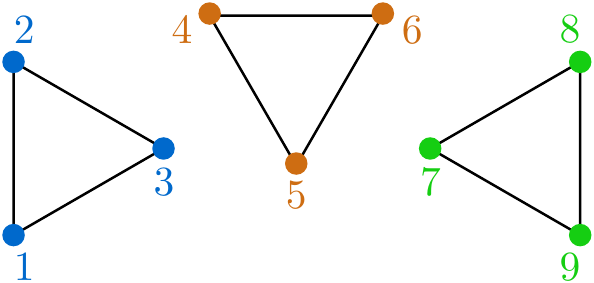}        
    \end{minipage}
	\hspace{0.04\linewidth}
    \begin{minipage}{0.44\linewidth}
	\includegraphics[width=\linewidth]{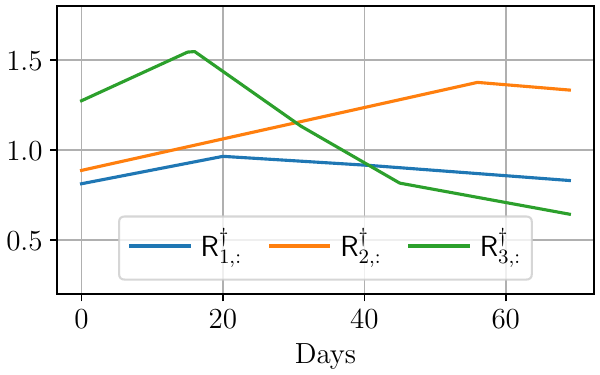}      
    \end{minipage}
	\includegraphics[width=\linewidth]{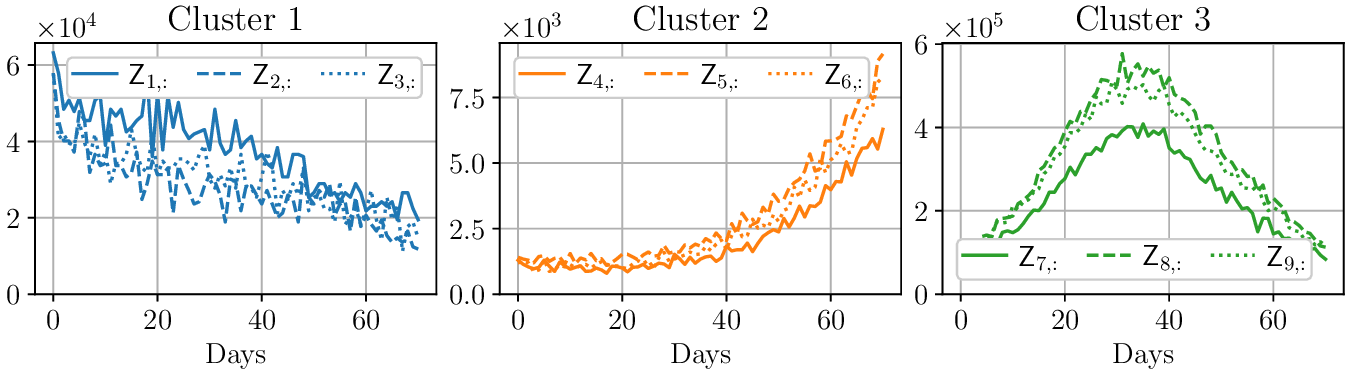}
	\caption{\textbf{Spatially structured synthetic data.} \textit{Top left:} Graph with $C = 9$ vertices split into $I = 3$ clusters; \textit{top right:} $I = 3$ realistic reproduction numbers constructed following~\cite{du2023compared,du2024synthetic}; \textit{bottom:} synthetic multivariate scaled Poisson counts in each of the $I=3$ clusters.}
	\label{fig:synthZ}
\end{figure}

The proposed $\joint$ estimator~\eqref{eq:joint_optim_pb} is validated  through intensive numerical simulations on carefully designed realistic synthetic data and then illustrated on real multivariate COVID-19 counts.

\begin{table}
    \centering
    {
    \begin{tabular}{cccccc}
    \toprule
    $\ML$ & $\Epi$ & $\fixL_\emptyset$ & $\fixL_\mathsf{b}$ & 
    $\fixL^{\star}$ & $\joint$ (ours) \\ \midrule
    $113.2 (\pm 4.5)$ & $27.7 (\pm 1.2)$ & $9.0 (\pm 1.1)$ & $5.4 (\pm 0.6)$ & $\mathbf{2.6 (\pm 0.4)}$ & $\mathbf{2.7 (\pm 0.4)}$ \\
    \bottomrule
    \end{tabular}
    }
    \caption{\textbf{Reproduction number estimation accuracy.} Performance on synthetic data following Fig.~\ref{fig:synthZ} are quantified by $10^{-4} \times \mRSE$~\eqref{eq:mRSE}, averaged over 20 generations and accompanied by $95\%$ Gaussian confidence intervals.}
    \label{tab:results}
\end{table}
\noindent \textbf{Synthetic multivariate scaled Poisson counts.} 
Another major contribution of the present work consists in the design of realistic synthetic \emph{scaled} Poisson 
spatiotemporal counts following~\eqref{eq:multivar-Z} accompanied by ground truth reproduction numbers and a prescribed connectivity structure.
Indeed, the graph diffusion-based procedure to generate spatiotemporal synthetic data developed in~\cite{du2024synthetic} induces non negligible correlations between reproduction numbers in path-connected territories not directly connected by an edge, departing from the desired connectivity structure, and restricts to \emph{standard} Poisson data.
To strictly respect the prescribed spatial structure, only graphs composed of fully connected subgraphs disconnected from each other are considered. 
In practice,  the $C$ \emph{abstract} countries are partitioned into $I$ clusters $\mathcal{C}_1, \dots, \mathcal{C}_I$, constituting the prescribed spatial structure.
In the following experiments, $C=9$ abstract countries are considered and split into $I=3$ clusters, as illustrated in Fig.~\ref{fig:synthZ} (\textit{top left}).
Then, for each cluster $\mathcal{C}_i$,  one realistic COVID-19 reproduction number time series $\Rvect_{i, :}^\dagger \in \rset_+^{T}$ is generated following~\cite{du2023compared,du2024synthetic} and used as ground truth for all countries of the cluster: $\forall c \in \mathcal{C}_i, \,  \Rvect_{c, :}^\star = \Rvect_{i, :}^\dagger$, hence exactly enforcing the desired spatial structure.
Finally,  initial counts $\Zvect_{:,0}$ are fixed and scale parameters are chosen large enough to allow variability between counts inside each cluster via $\gamma_c = 0.01 \times \Z^0_c$.
Finally multivariate scaled Poisson synthetic counts $\Zvect_{c,:}$ are independently sampled,
following Eq.~\eqref{eq:multivar-Z}. 
Fig.~\ref{fig:synthZ} displays the realistic reproduction numbers (\textit{top right}) and a generation of synthetic counts (\textit{bottom}) for each of the three clusters.

\noindent \textbf{Experimental protocol.} 
Performance of the reproduction number estimators are measured through the \textit{mean Relative Squared Error}:
\begin{align}
\label{eq:mRSE}
\mRSE(\Rvect) = \frac{1}{CT}\sum_{c=1}^C\sum_{t=1}^T \left(\frac{\widehat{\R}_{c,t} - \R^\star_{c,t} }{\R^\star_{c,t}}\right)^2 \,. 
\end{align}
The proposed $\joint$  strategy is compared with five methods.
The $\ML$ estimator consists in maximizing the likelihood associated to~\eqref{eq:multivar-Z}, leading to  $\widehat{\R}_{c,t}^{\ML} = \Z_{c,t}/\Zphi_{c,t}$.
The $\Epi$ method~\cite{cori2013new,thompson2019improved} processes \emph{independently} infection counts of all countries in a Bayesian framework with a temporal smoothing on a window of length $\tau$.
To test the robustness of the fixed graph method $\fixL$~\eqref{eq:argmin_R} to errors in the prescribed spatial structure, three spatial structures encoded in the Laplacian matrix are used: an empty graph $\L_\emptyset$,  a \emph{blurred} version $\L_\mathsf{b}$ of the true graph, with spurious edges of small weights,  and the true graph $\L^{\star}$.
Regularization parameters  are optimized through grid searches and the best $\mRSE$ is reported.
Performance are averaged over $20$ independent synthetic counts generations 
and uncertainty is assessed through $95\%$ Gaussian confidence intervals.

\noindent \textbf{Algorithmic settings.}
To mimic real world data for which the scale parameters $\gamma_1, \hdots, \gamma_C$ are unknown,  the Kullback-Leibler weights in~\eqref{eq:argmin_R} are chosen following the heuristic $\forall c \in \lbrace 1, \hdots, C \rbrace, \, \wdkl_c = 1/\mathrm{std}\bigl( \Zvect_{c,:}\bigr)$ inspired from~\cite{abry2020spatial,pascal2022nonsmooth}.
In the iterative procedures computing the $\fixL$ and $\joint$ estimates, the reproduction numbers are initialized at the countrywise $\Epi$ estimates with a sliding window of $\tau = 7$~days.
The number of iterations in Algorithm~\ref{alg:AO} is set to $\italt_{\max} = 10$.
Regularization parameters are optimized through exhaustive search across: all odd numbers between $1$ and $29$ for the sliding window length $\tau$ in $\Epi$ method; logarithmic grids of $\lambdatime$, $\lambdaspace$ and $\lambdafro$ ranging respectively from $1$ to $100$, $0.01$ to $1000$ and $0.001$ to $100$ for the penalized estimates, exploring $64$ values of $\lambdatime$ for $\fixL_\emptyset$,  a grid of $16\times16$ couples $(\lambdatime,\lambdaspace)$ for $\fixL$ and a grid of $8 \times 8 \times 8$ triplets $(\lambdatime, \lambdaspace,\lambdafro)$ for the $\joint$ estimator.

\noindent \textbf{Discussion.}
Table~\ref{tab:results} reports estimation accuracies achieved by the six considered methods.
Both the naive $\ML$ (first column) and state-of-the-art $\Epi$~\cite{cori2013new,thompson2019improved} (second column) lie far behind the penalized $\fixL$ and $\joint$  strategies (third to sixth columns) in terms of $\mRSE$.
The $\fixL$ method is robust to mild mismatches between the prescribed spatial structure used in $\fixL$ and the true connectivity graph (fourth and fifth columns).
Though, the $\fixL_\emptyset$ estimator enforcing only temporal regularity (third column) is superseded by the  $\fixL_\mathsf{b}$, $\fixL^{\star}$ and $\joint$ (fourth to sixth column) strategies including spatial regularization.
Finally, the $\joint$ method (sixth column) reaches the same excellent accuracy as the $\fixL$ one (fifth column) using the true graph.
Indeed, the estimated connectivity structure is almost exact, achieving a relative squared error of $\lVert\widehat{\L}^{\joint} - \L^{\star}\rVert_{\Fro}^2 / \lVert\L^{\star}\rVert_{\Fro}^2 = 2.3 (\pm 1.0) \times 10^{-13}$ and a perfect estimation of the support of $\L^{\star}$ when thresholding weights at $10^{-14}$.

\begin{figure}
    \centering
    \includegraphics[width=0.55\linewidth]{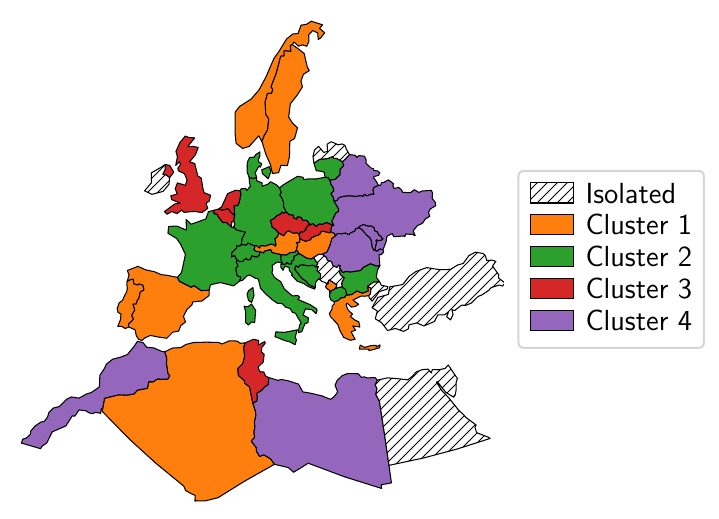}
    \includegraphics[width=0.44\linewidth]{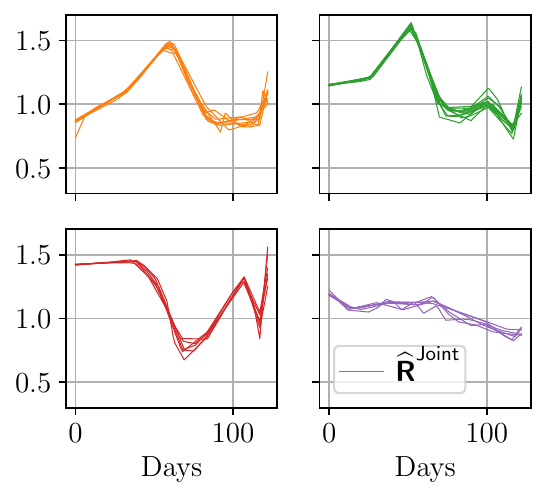}
    \caption{\textit{Left}: Map of the countries from the real data experiments. Colors indicate clusters derived from the spatial structure infered by $\joint$. Hashed countries are isolated ones. \textit{Right}: Reproduction numbers estimated from the $\joint$ method organized by clusters.  }
    \label{fig:map}
\end{figure}

\noindent \textbf{COVID-19 data.} 
The $\joint$ method is illustrated on a selection of real COVID-19 infection counts in $39$ countries in Europe and Africa between Sept.  $1$, $2020$ and Oct. $1$, $2021$
downloaded from Johns Hopkins University's repository.\footref{ft:jhu} 
The optimal hyperparameters estimated through grid search on synthetic data are manually fine-tuned to adjust the regularization level.\footnote{\label{ft:realexpelink} A \href{https://github.com/elasalle/EpiJointSpatiotempEstim/blob/main/experiments/real_data/europe_africa_expe.ipynb}{notebook} provides precise experimental setup and extra results.} 
The inferred graph shows four exactly disconnected clusters (in red, orange, green and purple) and a few isolated countries (dashed regions), displayed in Fig.~\ref{fig:map} (\textit{left}).
The estimated reproduction numbers in each country are grouped by cluster in Fig.~\ref{fig:map} (\textit{right}).\footref{ft:realexpelink} 
The $\joint$ method captures the main representative epidemiological dynamics across a large geographical area, which would have not been possible through visual inspection of the low quality counts, examples of which are displayed in Fig.~\ref{fig:trueZs}.
Interestingly, focusing on epidemiological dynamics,  the $\joint$ method does not only group together geographically close countries, \eg{,}  United-Kingdom, Tunisia and Czech Republic in the red cluster.

\section{Conclusions and perspectives}
\label{sec:conclusion}

A joint estimator of reproduction numbers and spatial connectivity structure has been proposed.
Intensive numerical experiments demonstrated its remarkable ability to infer spatiotemporal epidemiological indicators in an accurate and robust manner, while providing precise insights on the underlying spatial pattern.
The robustness of this $\joint$ method to change in the pool of monitored territories and in the phase of the pandemic will be systematically explored on real data.
For reproducibility, the code is publicly available on  Github.\footnote{\href{https://github.com/elasalle/EpiJointSpatiotempEstim}{github.com/elasalle/EpiJointSpatiotempEstim}}

\FloatBarrier

\bibliographystyle{unsrtnat}
\bibliography{refs.bib}
\end{document}